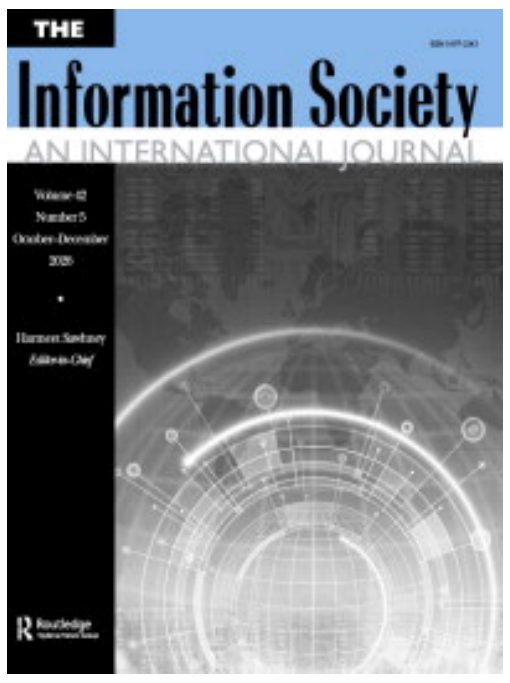

# The question of AI authorship: Literary-theoretical perspectives

Cagdas Duman, Imke van Heerden & Anil Bas

# The question of AI authorship: Literary-theoretical perspectives

Cagdas Duman[a,b], Imke van Heerden[c,d] and Anil Bas[e,f]

[a]Department of Communication Studies, Hacettepe University, Ankara, Turkey; [b]Department of Western Languages and Literatures, Bogazici University, Istanbul, Turkey; [c]Department of Digital Humanities, King's College London, London, UK; [d]Department of Comparative Literature, Koç University, Istanbul, Turkey; [e]National Centre for Computer Animation, Bournemouth University, Poole, UK; [f]Department of Computer Engineering, Faculty of Technology, Marmara University, Istanbul, Turkey

**ABSTRACT**
In this article, we focus on the conceptualization of Artificial Intelligence (AI) as a literary author. Although it has received consideration in other disciplines, the conceptual issues remain comparatively underexplored in literary studies. In an era that challenges the hegemony of the human, how should we posit AI as author? What possibilities, or impossibilities, could be offered by way of literary approaches to generative AI? Our exploration engages with an array of theoretical frameworks on authorship, text and textuality, readership, and interpretation. We revisit critical perspectives from the first and second halves of the twentieth century as well as more contemporary turns in literary theory to reflect on computer-generated text and ensuing interpretive methods. In doing so, we place the rethinking of authorship within a history of thought from which AI scholarship can still learn.

"Will all texts in the future be authorless and
nameless, written by machines for machines?
Is the future of literature reducible to mere code?"
—Kenneth Goldsmith (2011), *Uncreative Writing: Managing Language in the Digital Age*

"The author is dead and *has a new name*
*What it can be*
*What it was created to be*
*An alternate, more accurate rendition*"
—*GPT-2*[1]

## Introduction

AI systems have long demonstrated the capacity to generate a variety of creative texts: the scripts of the science-fiction short film *Sunspring* (Newitz 2016) and an imitation episode[2] of *Scrubs* (Britton 2017), a Japanese short story[3] (Sato 2016), a Harry Potter fan fiction chapter (Liao 2017), as well as Shakespearean sonnets (Lau et al. 2018) and Dr Seuss-styled poetry (Canales 2020). Textual quality has continued to increase, especially following the release of OpenAI's language model GPT-3 (Brown et al. 2020), as its op-ed in *The Guardian* illustrates: "Critics hope to refute what they consider as being the naivety of my voice. Yet there is more here than meets the eye!" (GPT-3 2020).

How is the quality of these texts evaluated? Early work in the field of creative text generation, especially AI poetry generation, established human evaluation as a typical practice for confirming quality. Machine-generated and human-written texts are jointly and anonymously presented to human judges, tasked with determining human likeness following predetermined evaluation criteria (Hopkins and Kiela 2017; Jhamtani et al. 2019; Lau et al. 2018; Solaiman et al. 2019). These texts are thus evaluated comparatively, with quality equated to humanness. The findings of Hopkins and Kiela (2017), for example, incidentally reveal participants' assumptions about what makes writing human, with half of the evaluators unable to identify texts by Dickinson, Dryden, Tennyson, and Shakespeare as human written (Van Heerden and Bas 2021, 180). However, in an era that questions anthropocentrism and in which generative AI complicates conventional notions of authorship, should the need to differentiate between machine- and human-created texts be problematized? Relatedly, given the current preoccupation with AI detection, including textual analysis and watermarking, to what degree is such differentiation possible, technically and intrinsically? There is consensus in publishing that a language model cannot be credited as an author since it cannot be held

**CONTACT** Anil Bas abas@bournemouth.ac.uk National Centre for Computer Animation, Bournemouth University, Poole, BH12 5BB, UK.

responsible for the work (COPE 2023), in addition to urgent moral and aesthetic concerns. The human lending their name to the piece takes ownership of and accountability for its content and purpose. The prevalent practice of detection, however, is more interested in the validity and trustworthiness of such attribution, treating AI disclosure, or lack thereof, as unreliable. These incongruities feed into a larger impasse surrounding authorship and provenance. The resolution, for some, lies in claiming generated output as theirs – the exact behavior that regulation seeks to prevent.

In this climate, we recognize the need for broad theoretical engagement to open new perspectives on this challenging issue. Drawing on literary and critical theory, we consider AI authorship through its productive tensions and ambiguities rather than through affirmation or critique. The point is not to make a narrow argument concerning formal attribution (e.g. adding the AI model's name as a co-author) or public disclosure of generative AI in a given piece of writing, but to engage more holistically with frameworks of *authorship*. The reality is that generative AI systems are being used, to varying degrees and in various ways, to *author* texts, which invites consideration. Therefore, we ask under what conditions one could apply the terms *author* and *literature* to a machine and its output, and do so whilst exploring related notions of text, textuality, and readership.

Although the past years have been particularly momentous for Large Language Models (LLMs), including ChatGPT, Claude, DeepSeek, Gemini, Grok, Kimi, and Qwen, possible connections between natural language processing research and literary-theoretical perspectives have remained largely unexplored. Discussions on the subject of AI authorship and natural language generation have taken place in domains such as computational creativity (Colton and Wiggins 2012; Gervás 2009), digital humanities (Van der Weel 2015), creative writing (Tucker 2019), media studies (Montal and Reich 2017; Tandoc Jr., Yao, and Wu 2020), philosophy of science (Du Sautoy 2019; Miller 2019), ethics (Hauer 2022; Orr and Davis 2020; Shank, DeSanti, and Maninger 2019), and legal studies (Bently 1994; Bridy 2012, 2015; Craig and Kerr 2021; De Vries 2020; Gervais 2020; Ruipérez et al. 2017). In this regard, legal scholarship concentrates predominantly on copyright and intellectual property laws – some of which do incorporate literary insights. Bently (1994) points to a variance in the conception of authorship between copyright law, which draws on Romantic authorship, and poststructuralist theories of literature. Bridy (2012) states that "the figure of the Romantic author sits monolithically at the core of copyright law, obscuring important realities about the collective nature of creativity and misrepresenting the actual processes of cultural production – both past and present" (4). In effect, creativity must be regarded in a more overarching fashion beyond the constrictions of the Romantic author. Other disciplines approach the question of the synthetic author from different angles. Media studies, for instance, explores the question of authorial attribution (Montal and Reich 2017) and "the impact of declared authorship (human journalist, an algorithm, or combined) and objectivity on news credibility perceptions" (Tandoc Jr., Yao, and Wu 2020, 549). In their interviews with industry experts, Montal and Reich (2017) identify a shared "humanized and anthropomorphic perception of authorship" (837). Somewhat differently, digital humanities revises the status of authorship, acknowledging "the birth of the new digital text technology" and examining "how the digital paradigm is likely to affect our thinking" (Van der Weel 2015, 2, 10). For Van der Weel (2015), these technologies necessitate new definitions of the author.

Scholarship bringing literary studies and AI into conversation includes, but is not limited to, Aquilina (2017), Elstermann (2020), Van de Cruys (2020), Henrickson (2021), Tencent Research Institute (2021), Van Heerden and Bas (2021, 2024), Neuman and Danesi (2022), Montfort and Heflin (2024), Floridi (2025), and Slocombe and Liveley (2025). The novelty of our work lies in engaging directly with questions concerning the AI author with the explicit aim of integrating literary theory into sociotechnical research, extending beyond the field's boundaries to place relevant insights in conversation with topical issues. To do so, we apply critical perspectives, using literary theory as a springboard to conceive authorship in light of AI technologies, touching on key concepts such as *authorial intent*, *author-function*, *intertextuality*, and *deconstruction*. In this manner, we address the challenge that AI poses to writing and creative practice.

Note that we trace the development of major theoretical frameworks largely in the order in which they emerged. In the subsequent discussions, however, we do not always strictly adhere to the original thinking. Rather, we place theoretical reflections in conversation with the contemporary phenomenon of AI authorship. In other words, the scholarly texts we discuss are approached through the lens of generative AI, inevitably conditioning how we engage with and interpret them. In the first section of this article, we use Calvino's ideas as a stepping stone to explore the author as a writing machine. In the second section, we concentrate on modernist, new critical, structuralist, and poststructuralist approaches to the author and authorship, referring to and extrapolating from Barthes, Derrida, Eco, Eliot, Foucault, Gallop, and Wimsatt and Beardsley. In the third section, we examine Deleuze's and Felski's theories as well as Cave's work on intelligence. In the fourth section, we utilize Haraway's concept of kin-making to exemplify posthuman stances. We conclude by highlighting how literary perspectives can inform the reconceptualization of authorship that AI necessitates.

## The author as writing machine

First presented in a 1967 lecture, the Italian writer Italo Calvino (1986) suggested that the writer has always been a kind of machine, rearranging words, rules, and conventions that already exist: "The 'I' of the author is dissolved in the writing. The so-called personality of the writer exists within the very act of writing: it is the product and the instrument of the writing process. … Writers, as they have always been up to now, are already writing machines" (15). The author contends in a prescient provocation that "my place could perfectly well be occupied by a mechanical device" (Calvino 1986, 16–17). Certainly, a real human being or "I" exists, who paints the canvas, pens the page and, more recently, prompts the AI model. This writer, on Calvino's account, is such a machine, a processor of language. Furthermore, Calvino calls the writer's personality a *product* of the writing. In a contemporary interpretation, this could mean that the persona of the author is created via the text and may or may not correspond with the identity of the flesh-and-blood writer. These ideas are explored further in the following discussion, in which the author as a construct (or, more simply, as a concept) is an idea important to understanding frameworks of authorship.

## Perspectives on authorship: Modernism, new criticism, structuralism, and poststructuralism

Indeed, Calvino's ideas resonate with those of major critics, philosophers, and literary figures of the twentieth century. A notable example is T.S. Eliot and modernist poetics, in particular, his theory of poetic impersonality. In the prominent essay "Tradition and the Individual Talent," Eliot (1919) opposes the Romantic conceptualization of the author, arguing that "[t]he progress of an artist is a continual self-sacrifice, a continual extinction of personality" (55). Eliot's essay shifted critical focus from the biography of the poet to the poetry. In fact, Eliot argues that the poet must escape their personality, cultivating a creative mind capable of absorbing and transforming a range of feelings and experiences. This is a striking point, given how contemporary art and literature are popularly framed as personal expression, as threads connecting readers to writers. Eliot argues that the poem must rather be situated in relation to existing poems. In doing so, he discusses the surprisingly dynamic concept of tradition. When truly novel poetry appears, it influences this tradition, altering the relations between existing texts. Poetic tradition is perpetually in motion in the sense that the depersonalized poet constantly re-explores and re-defines it. Eliot's traditionality and poetic depersonalization align with AI in interesting ways. An AI system is already apersonal in that it lacks personhood. Furthermore, AI is often portrayed as a platform or library of the world's texts. But how can we articulate the difference between a poet's engagement with literary texts, and an LLM's training on the same body of work? To ask an even more theoretical question: to what extent does AI-generated text enter the literary tradition and affect the relations between texts? Does it change our relationships with the history of literature?

Returning to Calvino, the question of authorship takes a different but related form when he describes not only the writer but also language itself as a machine: "[m]ankind is beginning to understand how to dismantle and reassemble the most complex and unpredictable of all its machines: language" (Calvino 1986, 10). This could be compared to Saussure's influential structuralist approach, which treats language as a *system*. Saussure's work became a crucial part of the linguistic turn, paving the way for poststructuralism and influencing theorists such as Althusser, Barthes, Derrida, Kristeva, Lacan, and Lévi-Strauss (Leitch et al. 2010).

In the highly influential *Course in General Linguistics*, first published in 1916, Saussure (1959) expounds that language is a self-contained "system of signs in which the only essential thing is the union of meanings and sound-images" (15). Each of these signs "was to be seen as being made up of a 'signifier' (a sound-image, or its graphic equivalent), and a 'signified' (the concept or meaning)" (Eagleton 1983, 84). Moreover, Saussure (1959) states that "[t]he bond between the signifier and the signified is arbitrary" (67). Saussure, by signaling this arbitrariness, emphasizes that there is "no natural connection" (69) between *the signifier* and *the signified*. As he suggests, "in language there are only differences" (120).

Saussure's structuralism betokened a major shift, which is to say that "[m]eaning was neither a private experience nor a divinely ordained occurrence: it was the product of certain shared systems of signification" (Eagleton 1983, 93). As Culler (1975) suggests in *Structuralist Poetics*, structuralism concerns "the realization that if human actions or productions have a meaning there must be an underlying system of distinctions and conventions which makes this meaning possible" (4). Applying these views, we might say that language models learn and illustrate the self-contained linguistic system in which meaning-making processes are predicated on difference. Though not concerned with AI, Eagleton's (1983) explanation helps to clarify the relevance of this school of thought to the current discussion: "Saussure's stress on the arbitrary relation between sign and referent, word and thing, helped to detach the text from its surroundings and make of it an autonomous object" (99). In the context of AI, an independent system of signs is highly relevant to considering how output by language models can make sense to readers, when the model does not have access to material referents. As mentioned, the relation between the signifier and the signified is arbitrary. From Saussure's perspective, meaning arises from the differences between signs within the system. Departing from this theoretical claim is the view that the model's disembodiment or detachment from physical reality is a limitation. For example, in a prevalent argument, AGI (artificial general intelligence) would benefit from multimodal architectures integrating language with perceptual and embodied forms of data (Paolo, Gonzalez-Billandon, and Kégl 2024). Embodiment through robotics is one proposed route, reflecting the idea that a system could improve if it were able to "step outside," as it were. Saussure, however, could argue that language is inherently self-contained. It must be mentioned that his theory's foundational principles were later critiqued by poststructuralists. Language models, however, invite us to critically revisit Saussure and those influenced by his work. But what does this mean for our understanding of authorship? When the text is viewed as detached and autonomous, its origin (whether human or AI) and the conditions of its creation become less important. Meaning is thought to reside in the relations between elements, which can be learned and replicated, creating the impression of meaningful engagement, agency, and intent.

Halfway through the twentieth century, New Critical theorists Wimsatt and Beardsley (1946) problematized authorial intent in their influential essay "The Intentional Fallacy." It argues that "the design or intention of the author is neither available nor desirable as a standard for judging the success of a work of literary art" (468), and that "evaluation of the work of art remains public; the work is measured against something outside the author" (477). They stress the literary work's complex structures and public literariness whilst emphasizing its irreducibility to authorial intention. The literary work, then, lies beyond individualized – or indeed authorialized – intention. Reducing it to that intention occludes criticism. Would we learn (perhaps we may call this human learning) to read machine-generated texts without resorting to intention, which is "neither available nor desirable" anyway? Jakobson (1960) identifies six functions of language: emotive, referential, conative, metalingual, phatic, and poetic. The poetic function, which is particularly relevant here, "focus[es] on the message for its own sake" (Jakobson 1960, 356). Extending this, poetry foregrounds its own form rather than the poet's intention. We would like to add, that the idea is not to negate the writer's role in crafting the text. The interpretation of the text, however, is a different matter.

It is useful to mention Umberto Eco's (1989) influential text *The Open Work,* which was originally published in 1962, preceding deconstructive criticism. Eco argues that the meaning of the text is catalyzed by the reader, contending that this recipient of the text unfurls and furthers its meaning via a unique vantage point:

> … each individual addressee can refashion the original composition devised by the author. The addressee is bound to enter into an interplay of stimulus and response which depends on his unique capacity for sensitive reception of the piece. … Thus, his comprehension of the original artifact is always modified by his particular and individual perspective. In fact, the form of the work of art gains its aesthetic validity precisely in proportion to the number of different perspectives from which it can be viewed and understood. These give it a wealth of different resonances and echoes without impairing its original essence … (Eco 1989, 3)

Art, for Eco (1989), "is an 'open' situation, in movement. A work in progress" (23). Could AI-generated writing invite similarly diverse readings, eliciting interpretation rather than comparison with human writing? Could we learn to approach machine-generated texts as open works, with meaning that unfolds through reception rather than authorial intent or origin? In principle, yes: AI-generated writing may exemplify Eco's notion even more radically than human-authored texts. It compels the reader to supply interpretive coherence, positioning interpretation rather than authorship as the site of meaning-making. In other words, meaning does not arise from an identifiable authorial consciousness. Reading literature, then, is an active, subjective engagement with the text's structures of meaning.

With the advent of poststructuralism, what we encounter is a more complex emphasis on difference and textual uncertainty; the onus is on the reader/critic who announces the irretrievable loss of authorial agency. As Barry (1995) writes, with poststructuralism "we enter a universe of radical uncertainty, since we can have no access to any fixed landmark which is beyond linguistic processing, and hence we have no certain standard by which to measure anything" (61). Derrida's *deconstruction*, for instance, demonstrates "how texts come to embarrass their own ruling systems of logic," criticizes the centralization of "symptomatic points," and voluntarily enters into textual "*aporia*[s] or impasses of meaning, where texts get into trouble, come unstuck, offer to contradict themselves" (Eagleton 1983, 116). Textual meaning, then, is engendered through heterogeneous differentiations since language cannot escape from, what Derrida terms *différance*. Pointing to continual differing and deferring, it is "the play of differences" (Derrida 1973, 130). Deconstruction problematizes the very idea of foundations and origins of meaning, which, as Derrida accentuates, are made of and gesture toward endless differences and delays. They are without substance, center, and presence. Meaning, therefore, is unceasingly undermined because it is in a process of endless postponement. In *Of Grammatology*, Derrida (1997) also remarks that "the person writing is inscribed in a determined textual system" (160), so much so that, as he (in)famously states, "*[t]here is nothing outside of the text*" (158, italics in original). He elaborates that "what we call production is necessarily a text, the system of a writing and of a reading which we know is ordered around its own blind spot" (Derrida 1997, 164). In what ways might the AI author reflect this inherent inescapability of textual uncertainty and undecidability? If the words are irreducibly unstable, and the text is "ordered around its own blind spot," could a machine learning model take heed of the complex uncertainties of the text? In another influential book, *Dissemination*, Derrida asserts that

> [a] text is not a text unless it hides from the first comer, from the first glance, the law of its composition and the rules of its game. A text remains, moreover, forever imperceptible. Its law and its rules are not, however, harbored in the inaccessibility of a secret; it is simply that they can never be booked, in the *present*, into anything that could rigorously be called a perception. And hence, perpetually and essentially, they run the risk of being definitively lost. Who will ever know of such disappearances? (Derrida 1981, 63)

What would the AI author do with these disappearances, within this absence of presence? If a text is "forever imperceptible," can we expect an algorithm to perform on par with its human counterpart? This is a lost cause; a machine cannot measure up to its human counterpart, since that counterpart has always lacked originating authority. In "The Death of the Author" Barthes (1977) dismantles the authority of what he calls the "Author-God" (146), observing that

> writing is the destruction of every voice, of every point of origin. Writing is that neutral, composite, oblique space where our subject slips away, the negative where all identity is lost, starting with the very identity of the body writing … the voice loses its origin, the author enters into his own death, writing begins. (Barthes 1977, 142)

The beginning of writing, then, debunks the always already meager state of the author, showing that the author never held absolute authority over the text's meaning. Moreover, Barthes highlights that this is effectively nothing but a limitation on the text:

> Once the Author is removed, the claim to decipher a text becomes quite futile. To give a text an Author is to impose a limit on that text, to furnish it with a final signified, to close the writing. Such a conception suits criticism very well, the latter then allotting itself the important task of discovering the Author (or its hypostases: society, history, psyche, liberty) beneath the work: when the Author has been found, the text is "explained" – victory to the critic. Hence there is no surprise in the fact that, historically, the reign of the Author has also been that of the Critic, nor again in the fact that criticism (be it new) is today undermined along with the Author. (Barthes 1977, 147)

Barthes opposes authorial focality, opening the text instead to the necessary intrusion of the critic. He also uses the concept of *intertextuality*, first coined and examined by Kristeva. Drawing upon Saussurean *semiotics* and Bakhtinian *dialogism*, Kristeva (1986) uses the term to refer to the continual network of texts. Allen, in a sweeping account, defines it as follows:

> The systems, codes and traditions of other art forms and of culture in general are also crucial to the meaning of a work of literature. Texts, whether they be literary or non-literary, are viewed by modern theorists as lacking in any kind of independent meaning. They are what theorists now call intertextual. The act of reading, theorists claim, plunges us into a network of textual relations. To interpret a text, to discover its

> meaning, or meanings, is to trace those relations. Reading thus becomes a process of moving between texts. Meaning becomes something which exists between a text and all the other texts to which it refers and relates, moving out from the independent text into a network of textual relations. The text becomes the intertext. (Allen 2000, 1)

Barthes influentially dwells on intertextuality in "From Work to Text." He draws attention to textual heterogeneity and demonstrates that textual plurality gestures toward dissemination of meaning:

> The Text is plural. Which is not simply to say that it has several meanings, but that it accomplishes the very plural of meaning: an irreducible (and not merely an acceptable) plural. The Text is not a co-existence of meanings but a passage, an overcrossing; thus it answers not to an interpretation, even a liberal one, but to an explosion, a dissemination. (Barthes 1977, 159)

The Text, in other words, is resolutely multiplicitous and unencumbered by the fetters of singular meaning. Barthes further propounds:

> [t]he intertextual in which every text is held, it itself being the text-between of another text, is not to be confused with some origin of the text: to try to find the "sources", the "influences" of a work, is to fall in with the myth of filiation; the citations which go to make up a text are anonymous, untraceable, and yet *already read*: they are quotations without inverted commas. (Barthes 1977, 160)

Accordingly, the text is without origin and continually in conversation with other texts. Deliberating over these points, we might say that the AI author is already dead, for there is no originating consciousness behind its text. Meaning arises not through intention but through the critic's or reader's interpretive act. The question of whether AI-generated text can be read as literature rather than dismissed as "mere code" rests on its reception and extratextual framing, though intertextuality reveals how all writing, including AI's, may evoke other texts both deliberately and unintentionally. Positioned within textual pluralities and intertextual dimensions, AI-generated text participates in the same network of signification as human writing.

Barthes's understanding of "the death of the author" is also deployed in legal studies. In "The Death of the AI Author," Craig and Kerr (2021) criticize the longstanding Romantic notion of authorship through poststructuralist theoretical positions. The article calls into relief an "error about the ontology of authorship," arguing that legal studies should opt for a "relational" stance that renounces Romantic authorship for good (Craig and Kerr 2021, 42, 44). We would emphasize that a relational stance does not necessarily exclude AI, since relationality is not a unique human quality and can be explored beyond humanistic concerns. It is also possible to argue that creativity and authorship are qualities that do not belong to the human only. Literature provides us with an apt medium for such exploration, in which AI could play a potentially productive role.

Barthes's contention that the author is always already dead, that it is a stillborn subject and inefficacious agent, in addition to the intertextual dimensions of the text further attest to the phantasmagoric (non)presence of the author. This is suggested in a similar vein by Foucault (1984) in "What is an Author?," in which he writes that "[i]n writing, the point is not to manifest or exalt the act of writing, nor is it to pin a subject within language; it is, rather, a question of creating a space into which the writing subject constantly disappears" (102). He elucidates discursive possibilities which point toward, not an individualistic author per se, but the idea of *author* as a discursive function. In other words, the author "is an ideological product" (Foucault 1984, 119). That is, the author as function indicates a series of discursive practices. The *author-function* denotes an interstitial return of the author, as a "characteristic of the mode of existence, circulation, and functioning of certain discourses within a society" (Foucault 1984, 108). At the end of Foucault's essay, he presciently asks a set of crucial questions, which are also applicable to our concerns:

> All discourses, whatever their status, form, value, and whatever the treatment to which they will be subjected, would then develop in the anonymity of a murmur. We would no longer hear the questions that have been rehashed for so long: Who really spoke? Is it really he and not someone else? With what authenticity or originality? And what part of his deepest self did he express in his discourse? Instead, there would be other questions, like these: What are the modes of existence of this discourse? Where has it been used, how can it circulate, and who can appropriate it for himself? What are the places in it where there is room for possible subjects? Who can assume these various subject functions? And behind all these questions, we would hear hardly anything but the stirring of an indifference: What difference does it make who is speaking? (Foucault 1984, 119–20).

Is the AI author ideological? Are discursive practices made legible through texts generated by a machine? Since language models generate text by learning patterns from large-scale corpora, their outputs inevitably echo the cultural and ideological assumptions embedded in the training data. In computational contexts, this raises further questions about how algorithmic systems participate in and reproduce modes of discourse. It may be that AI authorship could be conceptualized in terms of language as commodity, which means recognizing that it reveals how language today is produced, circulated, and valued within technological and capitalist systems. In an alternative understanding, AI authorship disrupts language as a commodity by destabilizing the notion of the author as the source and owner of linguistic production. Considering language's structural normativities and capitalistic circulations in meaning-making processes, this would allow the algorithm to function as an ecstatic experience in and of itself, rather than an authorial placeholder. Or, indeed, we may briefly ask: "What difference does it make who is speaking?"

Gallop (2011), in *The Deaths of the Author: Reading and Writing in Time*, revisits the poststructuralist buzzword, delving into both "the abstract, polemical death of the slogan and a moving, more bodily death of the mortal author" (2). Moreover, the work seeks to find a temporal possibility in the author's desired, ghostly return, thus showcasing "how the literal and theoretical death of the author are entangled" (Gallop 2011, 24). She suggests that we cannot think of reading and writing without its relation to time. Gallop's attention to time also reminds us that writing is never static but continually unfolding beyond the moment of its production. Echoing this article's epigraph, the question arises as to who the reader of these texts might be, and what kinds of audiences they invite. Could we fathom an act of reading separate from the human?

## Rethinking AI through postcritical theory

Are we, therefore, no longer in territory governed by a central authorial figure? This section introduces interpretive methods that transcend hierarchical and human-centered positionalities. The text, in this light, denotes a vast potentiality that might be constrained by overly anthropocentric reading practices. Instead of prioritizing the human intellect, we may be responsive to how AI-generated text could potentially celebrate multiplicity. Practically, this gestures toward the need for a proliferation of generative models trained on more diverse linguistic and cultural materials. It is, furthermore, important to consider the cultural and historical conditions of textual production.

In "The Problem with Intelligence: Its Value-Laden History and the Future of AI," Cave (2020) assesses "the use of intelligence (and related concepts) to justify social, political, and economic hierarchies, both within and across societies" (29). According to Cave, "intelligence is a highly value-laden term that has been implicated in a matrix of domination since its rapid rise to widespread usage just over a century ago" (32). He argues that "critical evaluation of the value-ladenness of intelligence can and should inform debates about the ethics and impact of AI" (Cave 2020, 34). When designing AI systems, one should aspire to interrogate ingrained ideologies since "[m]uch research shows that it is much more likely to be those who are already marginalized who will be most affected by AI and data-driven technologies" (Cave 2020, 34). Scrutiny of the values embedded in the term *intelligence* should extend to the operations of AI systems. Contemporary debates on algorithmic bias reveal how language models that appear neutral or objective can reproduce entrenched hierarchies and exclusions. LLMs can do so, as Bender et al. (2021) and Weidinger et al. (2021) warn, by parroting the systemic prejudices and dominant narratives within their training datasets.

Does AI authorship make us more susceptible to, or aware of, the power structures that surround us and the violence mechanisms that plague us? Does it continue to reflect and reaffirm gendered, racial, and speciesist subjugations? Could AI-generated text diminish our critical faculties or inspire us to think and act? Is AI-generated writing political or completely de-politicized?

These points bring us to consider some postcritical positions and their precursors. Language is malleable, and imagined and (ab)used as a hierarchical system. Literary language carves out a pluralist space against standardized linguistic and intellectual violence. Language can be repurposed, then, to expose the limits of the human as well as how hierarchical structures are sustained. The former finds resonance with *Dialogues*, in which Deleuze and Parnet (1987) assert that "the aim of writing is to carry life to the state of a non-personal power. … Why does one write? Because it is not a case of writing. … To write has no other function: to be a flux which combines with other fluxes – all the minority-becomings of the world" (50). In *Essays Critical and Clinical*, Deleuze states:

> To write is certainly not to impose a form (of expression) on the matter of lived experience. Literature rather moves in the direction of the ill-formed or the incomplete … Writing is a question of becoming, always incomplete, always in the midst of being formed, and goes beyond the matter of any livable or lived experience. It is a process, that is, a passage of Life that

> traverses both the livable and the lived. Writing is inseparable from becoming: in writing, one becomes-woman, becomes-animal or vegetable, becomes-molecule to the point of becoming-imperceptible. (Deleuze 1997, 1)

Writing, for Deleuze, is the repudiation of identification; it is a disidentificatory process, a process of possibilities, a process of escaping, a process of becoming other than the author. It is also a desire to become indiscernible, to become irreducibly contiguous with other (ir)realities, to become interstitial, to be deterritorialized. Generative AI may epitomize this notion. Does AI authorship allow for new becomings? Does it, necessarily, further augment our (in)humanity? If understood under the banner of *authorship*, user interaction with generative AI unsettles anew the distinctions between human and machine, personal and impersonal, as well as author, text, and reader. This also invites reflection on the critical stance itself, i.e., how critics consider texts, technologies, and the act of interpretation.

More recently, Felski's trajectory-changing book, *The Limits of Critique*, questions longstanding interpretive methodologies:

> Why is it that critics are so quick off the mark to interrogate, unmask, expose, subvert, unravel, demystify, destabilize, take issue, and take umbrage? What sustains their assurance that a text is withholding something of vital importance, that their task is to ferret out what lies concealed in its recesses and margins? Why is critique so frequently feted as the most serious and scrupulous form of thought? What intellectual and imaginative alternatives does it overshadow, obscure, or overrule? And what are the costs of such ubiquitous criticality? (Felski 2015, 5)

Felski does not posit the text as something from which to deduce meaning by following supposedly hidden traces, but more like an entity with which to coevolve. She defines *postcritical reading* as "a matter of attaching, collating, negotiating, assembling – of forging links between things that were previously unconnected" (2015, 173). This is understood "as an act of composition – of creative remaking – that binds text and reader in ongoing struggles, translations, and negotiations" (Felski 2015, 182). Could AI authorship give rise to interpretive methodologies that exceed mere interrogatory analyses? Rather than treating the text as something to be decoded, one might give prominence to affective, aesthetic, and temporal engagement, emphasizing how meaning emerges with, and not from, the text. Such emphasis may set the stage for human-AI complementarity and, specifically, for co-creation between human and machine. As Van Heerden and Bas (2021) ask, could this kind of collaboration inspire new literary movements?

## Kin-making

Several of Haraway's works advocate for non-anthropocentric and nonhierarchical thinking. In *Staying with the Trouble: Making Kin in the Chthulucene*, she writes that "[n]o species, not even our own arrogant one pretending to be good individuals in so-called modern Western scripts, acts alone; assemblages of organic species and of abiotic actors make history, the evolutionary kind and the other kinds too" (Haraway 2016, 100). She therefore urges us to *make kin*:

> My purpose is to make "kin" mean something other/more than entities tied by ancestry or genealogy. The gently defamiliarizing move might seem for a while to be just a mistake, but then (with luck) appear as correct all along. Kin making is making persons, not necessarily as individuals or as humans. (Haraway 2016, 101–2)

Such a lens not only destabilizes the Barthesian "Author-God" but also encourages us to attend to proliferation by means of kin making. Perhaps, instead of focusing on the radical differences between a language model and a human author, we might recognize continuities between them, even considering how their authorial activities are mutually generative. Tucker (2019), for instance, is interested in human-machine symbiosis in creative writing and translation, understood beyond anthropocentric ideals as "a dense ecosystem of simultaneous and multiple actions and authors" (7). The theatrical production *AI: When a Robot Writes a Play* (THEaiTRE 2021) similarly represents the script's composition as a form of human-machine collaboration (Van Heerden, Duman, and Bas 2023, 107).

Making kin within assemblages is to celebrate co-occurrence and to rethink uncertainties as positive prospects. It means to adopt diverse viewpoints that challenge anthropocentrism and recognize creativity as a shared, interdependent process. Focusing on AI art and intellectual property, Zeilinger (2021) pertinently stresses the former's capacity "to instantiate agency as relational, decentered and plural, and to approach creative practice as fundamentally dynamic and embedded" (14). Braidotti's (2019) *critical posthumanities* significantly diversifies the "posthuman knowing subject," which is "understood as a relational embodied and embedded, affective and accountable entity and not only as a transcendental consciousness" (31). In this sense, what may appear to indicate the end of human intellect could simultaneously herald a beginning.

## Conclusion

Beyond literary studies, the idea of the solitary creative genius continues to hold sway when in fact AI renders visible what many frameworks have long intimated:

that authorship is situated and relational. Topics for future research in this context include human-AI complementarity and co-creativity; questions of provenance, originality, and accountability; the role of the reader (as well as reviewer and critic); interpretation and intertextuality; as well as the shifting boundaries between tool and agent. The advent of LLMs urges the reconceptualization of authorship. Our intervention with this article is to stage that rethinking through a history of literary and philosophical paradigms from which AI scholarship can still learn. Renowned essays and books predating LLMs by decades describe language as a self-contained system, the writer as a machine, or poetry as impersonal. The notion of the author as a product (of writing or of ideology) is a theoretical position that could inform attempts to frame AI authorship in imaginative ways. Generative AI's growing influence has led to a resurgence in authorial intention (including emphasis on the writer's agency, creativity, craft, and emotion). This could be interpreted as an attempt to articulate the humanness of human writing in the age of AI. Certain schools of literary theory, however, would suggest that a writer does not express a preexisting identity. Instead, the persona of the writer comes into being with the writing. This observation serves as a reminder that identities and ideas are constituted through discourse; they do not exist prior to or outside of language. Contemporary concerns over the power and influence of LLMs can be understood in this light. They possess the capabilities to shape the ideas that shape our world.

Literary-theoretical perspectives invite contemplation of structures of meaning-making that exceed the individual. Literature is an apposite medium to explore the dispersion of meaning, intelligence, agency, and creative responsibility across linguistic, cultural, and computational systems. If authorship is thus shared, the question of whether AI supersedes the human author loses consequence in the philosophical sense. Unfortunately, the professional and economic dimensions of writing, as well as the lived experiences of creative communities, suggest a far more disquieting possibility.

## Notes

1. This text was produced using GPT-2, a predecessor of the models behind ChatGPT (OpenAI 2022). The prompt "The author is dead and" was provided by the user and the italicized text was generated by the language model.
2. In October 2017, Botnik Studios' Twitter (now X) account posted a computer-generated script based on scripts of the well-known television series *Scrubs* (Britton 2017).
3. Sato (2016) submitted a computer-generated story entitled "The Day a Computer Writes a Novel" for the Hoshi Shinichi Award, a Japanese literary award that accepts both human-created and nonhuman-created texts limited to 10,000 characters.

## Disclosure statement

No potential conflict of interest was reported by the author(s).

## Author contributions

CRediT: **Cagdas Duman**: Investigation, Methodology, Writing – original draft; **Imke van Heerden**: Conceptualization, Funding acquisition, Methodology, Project administration, Supervision, Writing – original draft, Writing – review & editing; **Anil Bas**: Methodology, Project administration, Supervision, Writing – original draft, Writing – review & editing.

## Funding

This paper has been produced benefiting from the 2232 International Fellowship for Outstanding Researchers Program of TÜBİTAK (Project No: 118C285). However, the entire responsibility of the paper belongs to the owner of the paper. The financial support received from TÜBİTAK does not mean that the content of the publication is approved in a scientific sense by TÜBİTAK. Imke van Heerden is supported by the Engineering and Physical Sciences Research Council [grant number EP/Y009800/1], through funding from Responsible AI UK. Anil Bas has received funding from the EU H2020 Marie Skłodowska-Curie COFUND Scheme (Grant No. 900025), CfACTs.

## ORCID

Cagdas Duman http://orcid.org/0000-0002-8330-2965
Imke van Heerden http://orcid.org/0000-0002-4224-8800
Anil Bas http://orcid.org/0000-0002-3833-6023